\documentclass[preprint2]{aastex62}

\usepackage{amsmath}
\usepackage{color}

\shorttitle{\mosel\ survey: Size dependent quenching between $z=2-4$}
\shortauthors{Gupta et al.}

\newcommand{\mosel}{{\tt MOSEL}}

\newcommand{\oiii}{[\hbox{{\rm O}\kern 0.1em{\sc iii}}]}
\newcommand{\hb}{\hbox{{\rm H}\kern 0.1em{\sc $\beta$}}}
\newcommand{\msun}{M$_{\odot}$}
\newcommand{\logmstar}{$\log(M_*/{\rm M}_{\odot})$}

\newcommand{\exsitu}{{\it ex-situ}}
\newcommand{\rhalfmass}{$R_{\rm halfmass}$}

\begin{document}

\title{\mosel\ and IllustrisTNG:   Massive Extended Galaxies at $z=2$ Quench Later Than Normal-size Galaxies
}
\author[0000-0002-8984-3666]{Anshu Gupta}
\affiliation{School of Physics, University of New South Wales, Sydney, NSW 2052, Australia}
\affiliation{ARC Centre of Excellence for All Sky Astrophysics in 3 Dimensions (ASTRO 3D), Australia}

\author[0000-0001-9208-2143]{Kim-Vy Tran}
\affiliation{School of Physics, University of New South Wales, Sydney, NSW 2052, Australia}
\affiliation{ARC Centre of Excellence for All Sky Astrophysics in 3 Dimensions (ASTRO 3D), Australia}

\author[0000-0003-1065-9274]{Annalisa Pillepich}
\affiliation{Max-Planck-Institut f\"{u}r Astronomie, K\"{o}nigstuhl 17, 69117 Heidelberg, Germany}

\author[0000-0002-9211-3277]{Tiantian Yuan}
\affiliation{Swinburne University of Technology, Hawthorn, VIC 3122, Australia}
\affiliation{ARC Centre of Excellence for All Sky Astrophysics in 3 Dimensions (ASTRO 3D), Australia}

\author[0000-0001-9414-6382]{Anishya Harshan}
\affiliation{School of Physics, University of New South Wales, Sydney, NSW 2052, Australia}
\affiliation{ARC Centre of Excellence for All Sky Astrophysics in 3 Dimensions (ASTRO 3D), Australia}

\author[0000-0002-9495-0079]{Vicente Rodriguez-Gomez}
\affiliation{Instituto de Radioastronom\'ia y Astrof\'isica, Universidad Nacional Aut\'onoma de M\'exico, A.P. 72-3, 58089 Morelia, Mexico}

\author[0000-0002-3185-1540]{Shy Genel}
\affiliation{Center for Computational Astrophysics, Flatiron Institute, 162 Fifth Avenue, New York, NY 10010, USA}
\affiliation{Columbia Astrophysics Laboratory, Columbia University, 550 West 120th Street, New York, NY 10027, USA}

\begin{abstract}

Using the  TNG100 (100 Mpc)$^3$ simulation of the IllustrisTNG project, we demonstrate a strong connection between the onset of  star formation quenching and the stellar size of galaxies.  We do so by tracking the evolutionary history of extended and normal-size galaxies selected at $z=2$ with \logmstar$ = 10.2 - 11$ and stellar-half-mass-radii above and within 1-sigma of the stellar size--stellar mass relation, respectively. We match the  stellar mass and star formation rate distributions of the two populations. By $z=1$, only 36\% of the extended massive galaxies have quenched, in contrast to a quenched fraction of 69\% for the normal-size massive galaxies. We find that normal-size massive galaxies build up their central stellar mass without a significant increase in their stellar size between $z=2-4$, whereas the stellar size of the extended massive galaxies almost doubles in the same time.  In IllustrisTNG, lower black hole masses and weaker kinetic-mode feedback  appears to be responsible for the delayed quenching of star formation in the extended massive galaxies. We show that relatively gas-poor mergers may be responsible for the lower central stellar density and weaker supermassive black hole feedback in the extended massive galaxies.  %

\end{abstract}


\section{Introduction}

How and when massive galaxies stop forming stars (aka quenching) remains a mystery  after decades of study. A  significant fraction of massive galaxies have either stopped forming stars or show suppressed star formation in the first 15-20\% of the age of the Universe ($z=2-3$) \citep{vanDokkum2008, Damjanov2009, Straatman2014, Glazebrook2017}. 
The emerging picture is that star formation is quenched  in field galaxies either via feedback processes such as those related to supernovae and active galactic nuclei (AGN), or a truncation of gas supply to fuel the star formation \citep{Springel2005a, Zhang2019a}. In massive dark matter halos, the infalling gas can be shock-heated, further preventing the inflow of gas at virial radii \citep{Dekel2006}. At higher redshifts, mergers and violent disk instabilities can also trigger a burst of star formation and AGN activity in the core, subsequently quenching  star formation \citep{Dekel2014, Zolotov2015}. AGNs are thought to be  primarily responsible for the quenching of star formation in massive galaxies \citep{Fabian2012, Harrison2017a}.

The physical processes responsible for  quenching leave imprints on the spatial distribution of stars and stellar ages \citep{McDermid2015, Wu2018a}. A central starburst will rapidly quench star formation in the center, either by rapid consumption of gas or by triggering strong outflows, and produce galaxies with a dense stellar core \citep{Springel2005a, Snyder2011, Zolotov2015}. Less centrally focused feedback processes such as shock heating of gas in massive dark matter halos, energy injected by the supernovae and/or AGN   would quench star formation gradually without significantly altering the stellar distribution of galaxies \citep{Dekel2006, Fabian2012, Terrazas2017}. Therefore, we expect to see a correlation between the star formation histories and the spatial distribution of stars. 

Large extragalactic surveys in the local and high redshift Universe find that star formation in compact galaxies quench on a shorter time-scale compared with larger,  more extended, galaxies. In the local Universe, at a fixed stellar mass, compact galaxies have relatively older stellar populations indicative of either rapid star formation quenching and/or an earlier formation epoch \citep{McDermid2015, Barone2020}. Massive diffuse galaxies in the local universe such as Malin I build up their stellar disk slowly over an extended timescale \citep{Boissier2016, Zhu2018}.  Using data from the Sloan Digital Sky Survey, \cite{Fang2013} most galaxies quench after they cross a certain threshold in the central stellar mass density.

Studies at high redshift ($z\sim 1$) also find that massive compact galaxies  have older stellar populations compared with their larger counterparts, suggesting that star formation quenched in compact galaxies earlier \citep{Williams2017, Fagioli2016}.  In contrast, \cite{Keating2015} find that compact massive galaxies at $z\sim1.5$ have younger stellar ages compared with a control population, suggesting that compactness is the result of a recent burst of star formation.  Observations by \cite{Barro2017} show that the connection between the central stellar mass surface density and the status of the star formation exists as early as at $z~2.5-3$. Using spectroscopic data from the Lega-C survey \citep{vanderWel2016, Straatman2018},  \cite{Wu2018a} shows that massive galaxies at $z\sim 1$ seem to follow two distinct star formation quenching pathways: (i) slow ($>1$ Gyr) quenching producing quiescent galaxies with large galactic disks and younger stellar populations, and (ii) fast ($<1$ Gyr) star-formation quenching producing post-starburst galaxies with various sizes depending on their formation history.

\begin{figure}
	\centering
	\tiny
	\includegraphics[scale=0.23, trim=0.0cm 0.0cm 0.0cm 0.0cm,clip=true]{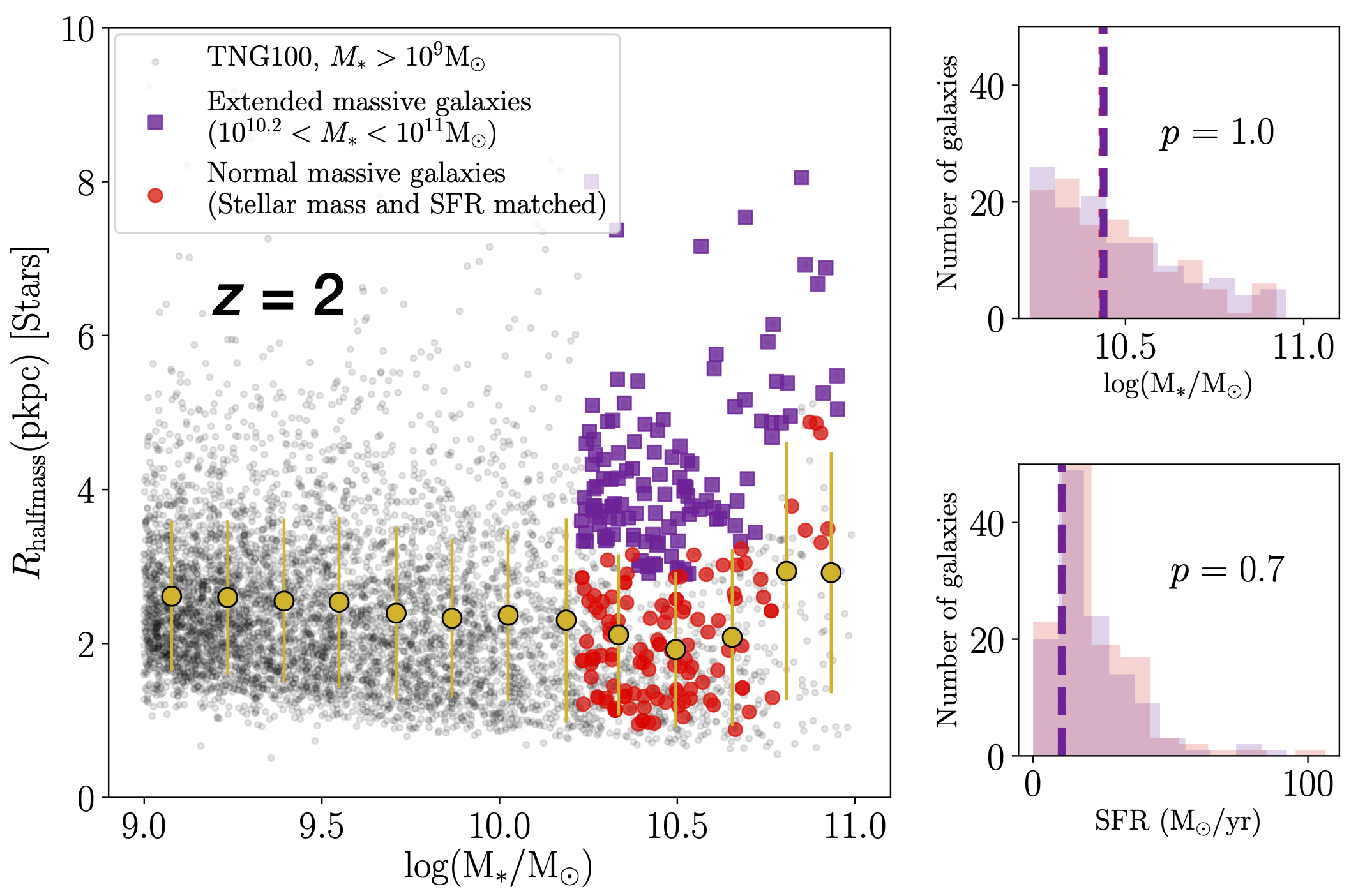}
	\caption{The stellar size--stellar mass relation for TNG100 galaxies at $z=2$. The golden circles represent average stellar sizes binned into equal stellar mass bins. The purple squares are the extended massive galaxies (\logmstar$\, = 10.2-11$) with stellar size 1-$\sigma$ above the mean size--mass relation. The red circles represent our control sample of normal-size (aka normal) massive galaxies. The two panels on the right show the stellar mass and SFR distributions of the  normal and extended massive galaxies at $z=2$. The dashed purple and dotted red line represent the median for the respective population and are statistically identical at $z=2$. We label the $p$-value from a two sample KS test in the respective panel.  }
	\label{fig:illustris_outliers_size}
\end{figure}

In cosmological simulations, massive compact galaxies at $z=2$  exhibit two main formation pathways: (i) a recent burst of star formation between $z=2-4$ triggered by gas-rich major merger or (ii) they formed at $z>4$ when the Universe was denser than today \citep{Wellons2015}. In agreement with observations, compact galaxies formed via mergers have relatively younger stellar populations, whereas compact galaxies formed at earlier epochs have older stellar populations. However, \cite{Wellons2015} focus on only a small fraction of extremely compact galaxies from the Illustris simulation with a stellar size almost 2-sigma below the galaxy size-mass relation and \logmstar\,$>11.0$.  

\cite{Genel2017} analysed the size evolution of galaxies in the IllustrisTNG simulations (\url{https://www.illustris-project.org}).
They find that most quenched galaxies at $z=0$ started off at the lower end of the stellar mass-size relations, i.e., quenched galaxies have relatively compact stellar distribution at higher redshift.  In \cite{Gupta2020}, we use the \mosel\ survey \citep{Tran2020} to track how galaxies at $z\sim3$ grow.  Combining our observations with predictions from IllustrisTNG, we show that massive galaxies after $z\sim3$ tend to form new stars from \exsitu\ processes. An intriguing result from \cite{Gupta2020} is a dramatic increase in the stellar size of massive galaxies at $z=2$.

In this follow-up study, we  investigate the size evolution of the massive galaxies and its link with the star formation quenching  by analysing the outcome of the publicly available IllustrisTNG simulations \citep{Nelson2019a}.
We select normal-size and extended massive simulated galaxies at $z=2$  according to their location on the stellar mass -- stellar size plane of IllustrisTNG and track their evolutionary histories backwards and forwards in time. To minimize the bias due to the different mass--size relations of quiescent and star-forming galaxies, also in place in the simulations \citep{Genel2017}, we select two samples of galaxies with consistent stellar mass and star formation rate distributions at the time of selection, here $z=2$.

\begin{figure*}
	\centering
	\tiny
	\includegraphics[scale=0.32, trim=0.0cm 0.0cm 0.0cm 0.0cm,clip=true]{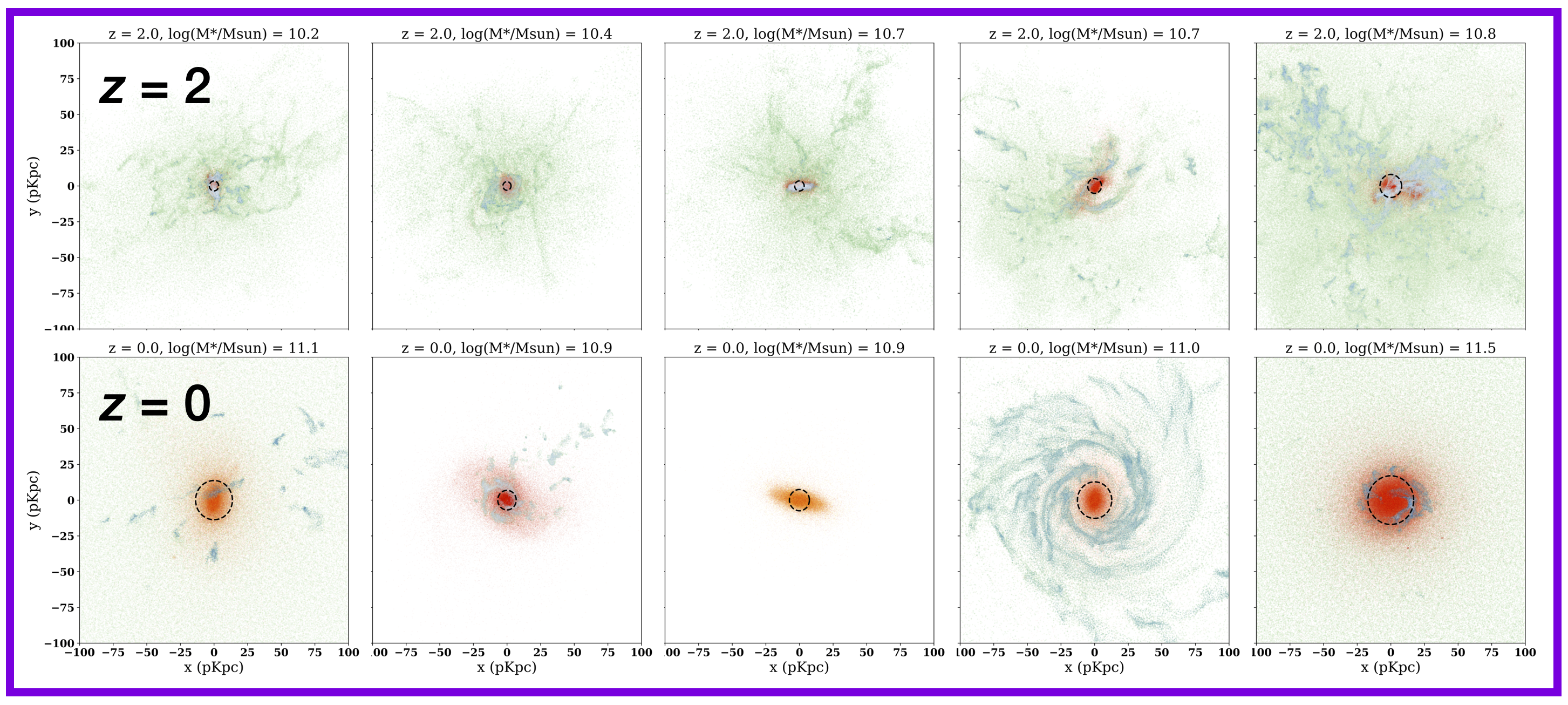}
	\includegraphics[scale=0.32, trim=0.0cm 0.0cm 0.0cm 0.0cm,clip=true]{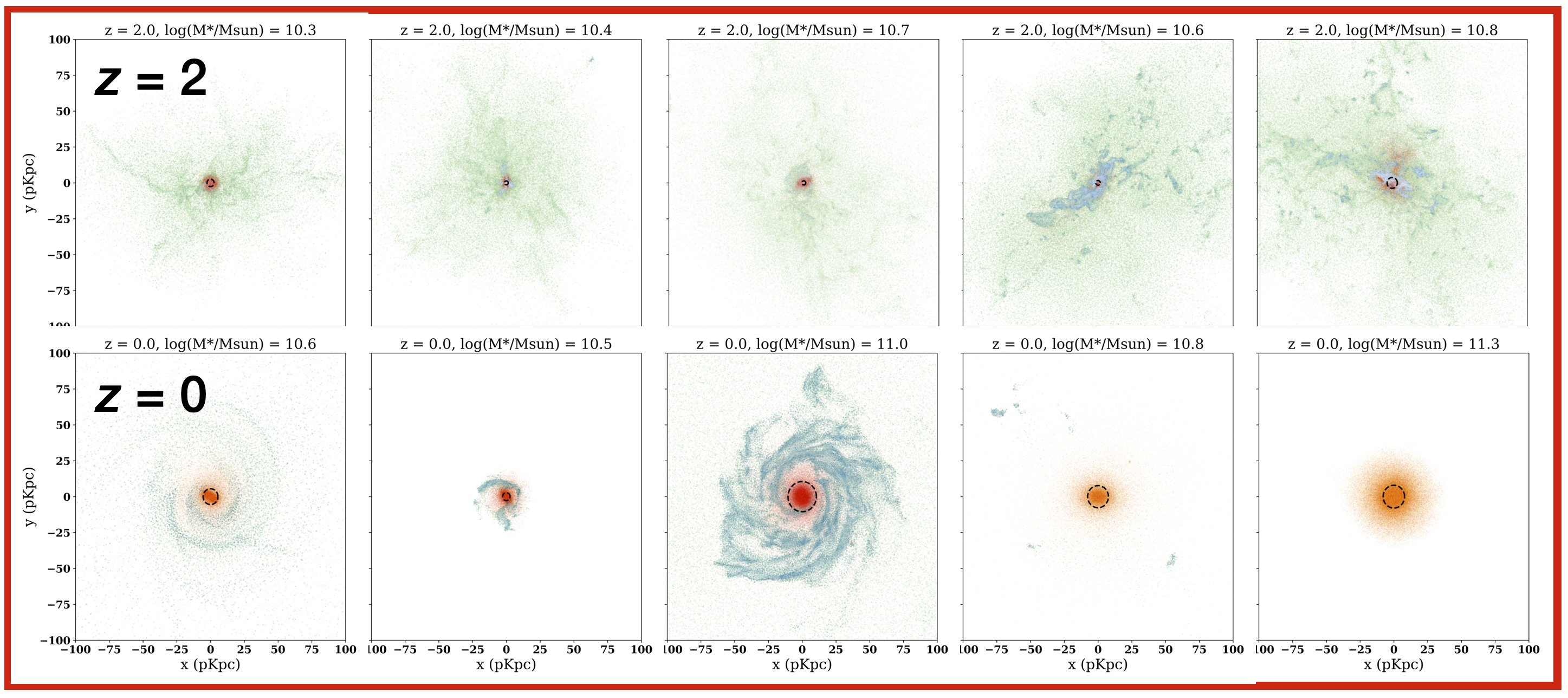}	
	\caption{ Distribution of stars (red), hot gas (green) and neutral gas (blue) in 5 randomly selected  TNG100 simulated galaxies in increasing order of stellar mass. From top to bottom: (i) top two rows: extended massive galaxies at $z=2$ (first) and their descendants at $z=0$ (second row); (ii) counterpart normal-size (aka normal) massive galaxies at $z=2$ (third row) and their descendants at $z=0$ (bottom row). The dashed circle in each panel corresponds to the stellar half mass radii.  }
	\label{fig:illustris_galaxies}
\end{figure*}

The paper is organised as follows. In Section \ref{sec:methods}, we provide a brief description of the simulations and our selection method of the extended and normal-size massive galaxies. In Section \ref{sec:results} we compare the star-formation histories of the two samples and how the two populations evolve on the size-mass plane.   In Section \ref{sec:discussion} and \ref{sec:conclusion}, we discuss and summarise our main findings.

\section{Methods} \label{sec:methods}
\subsection{The TNG100 simulation of IllustrisTNG}

We use a cosmological hydrodynamical simulation from the IllustrisTNG project \citep{Pillepich2017a, Nelson2017, Springel2017,  Marinacci2017, Naiman2017} to understand the relation between stellar disk size and star formation history of massive galaxies. In particular, we use the publicly available outcome \citep{Nelson2019a} of the TNG100 simulation ($\sim$[100 Mpc]$^3$, $m_b = 9.4 \times 10^5$/h) to ensure a statistically significant sample of galaxies with \logmstar$>9$ at $z\gtrsim3$, i.e., galaxies with $\sim$1000 baryonic resolution elements. We start from a selection of galaxies with \logmstar$>9$ at $z=0$ and track them back in time using the merger tree catalogs  \citep{Rodriguez-Gomez2015}. 

\begin{figure*}
	\centering
	\tiny
	\includegraphics[scale=0.45, trim=0.0cm 0.0cm 0.0cm 0.0cm,clip=true]{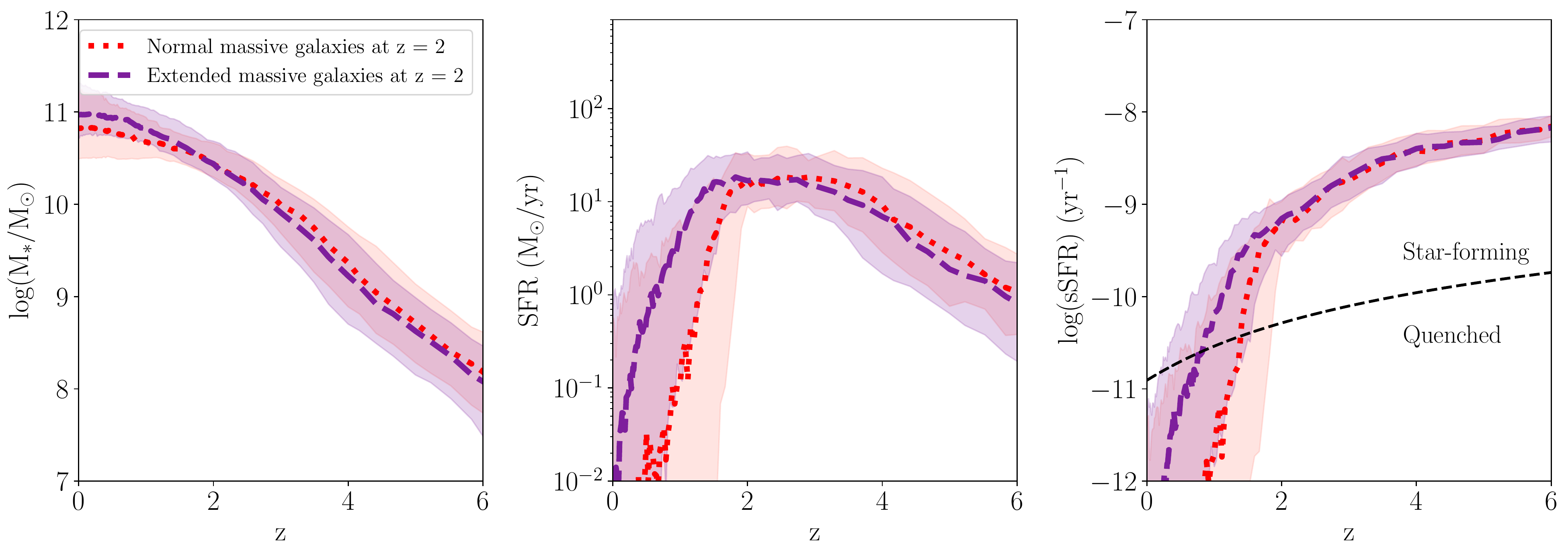}
	\caption{The redshift evolution of stellar mass (left), star formation rate (middle), and specific star formation rate (right) of the extended (purple) and normal-size (red) massive galaxies  from the TNG100 simulation. The dashed curves and shaded region show the median and 16$^{\rm th}-$84$^{\rm th}$ percentiles.  The star formation in the extended massive galaxies quench later than the normal-size massive galaxies. The dashed black curve corresponds to the redshift dependent sSFR cut-off for the quenched galaxies \citep{Tacchella2019}, which is adopted throughout to separate quenched versus star-forming galaxies.  }
	\label{fig:illustris_outliers_sfh}
	
\end{figure*}

\subsection{Selection of extended vs. normal-size massive galaxies}

To select extended massive galaxies in TNG100, we bin the  stellar mass--size relation at $z=2$ into bins of $0.15\,$dex in stellar mass. Throughout this paper, we use the three-dimensional  stellar half-mass-radius (\rhalfmass\ =  radius containing 50\% of the stellar particles) to represent the stellar size, or stellar extent, of a galaxy. All properties of galaxies are measured within twice the stellar half-mass-radius, unless otherwise specified. We select massive galaxies with stellar mass $10.2<$\logmstar$<11$ at $z=2$ with stellar sizes $1\sigma$ above the stellar mass-size relation (Figure \ref{fig:illustris_outliers_size}): this leads to the selection of 123 extended massive galaxies. We limit our extended massive galaxies sample to \logmstar$<11.0$ due to the low number of galaxies at higher stellar masses.   


 From TNG100, we build a control sample of normal-size simulated galaxies (normal galaxies hereafter). For each extended massive galaxy, we select a normal-size massive galaxy whose \rhalfmass\ is within $1\sigma$ of the median stellar mass--stellar size relation, and whose stellar mass and  star formation rate (SFR)  match the corresponding values of the extended massive galaxy within $0.1\,$dex at $z=2$.
This ensures statistically identical stellar mass ($p=1.0$) and SFR ($p=0.7$) distributions for the normal-size and extended massive galaxies at $z=2$ (Figure \ref{fig:illustris_outliers_size}: right-hand panels). This  
results in a minimal stellar mass and SFR bias between the two sample upon selection. Under this selection at $z=2$, the TNG100 extended massive galaxies have on average twice the stellar \rhalfmass\  compared with  the normal-size massive galaxies. 

Figure \ref{fig:illustris_galaxies} shows the distribution of gas and stars in five randomly selected extended and normal-size massive galaxies at redshift  $z=2$ and their descendants at $z=0$. It demonstrates the diversity of galaxy types in both populations. In the following sections, we track the evolutionary histories of extended and normal-size massive galaxies to understand the build up of the extended stellar disk and its connection with the star formation quenching or suppression. We use the merger tree catalogs by \cite{Rodriguez-Gomez2015} to identify progenitors and descendents of our extended and normal-size massive galaxies. 

\begin{figure*}
	\centering
	\tiny
	\includegraphics[scale=0.45, trim=0.0cm 0.0cm 0.0cm 0.0cm,clip=true]{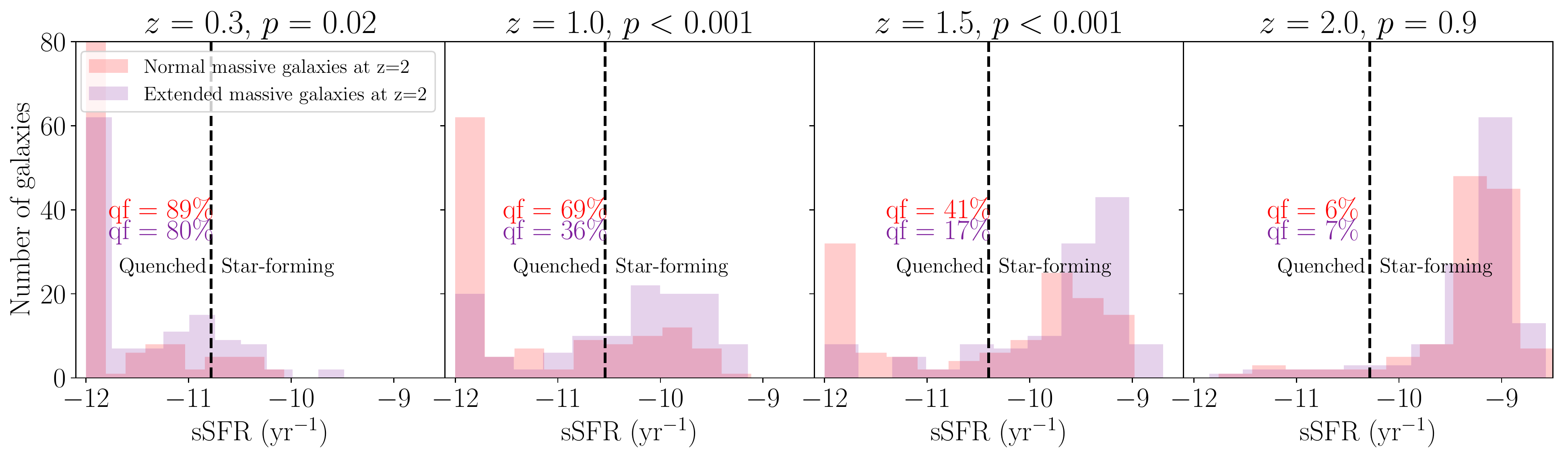}
	\caption{The specific star formation rates of the extended (purple) and normal (red) massive galaxies selected in TNG100 at $z=2$ and here shown at four redshifts: $z=0.3, 1.0, 1.5$, and $2.0$, from left to right.  In each panel, we label the quenched fraction ($qf$) for the respective  population, calculated based on the dashed black line in Figure \ref{fig:illustris_outliers_sfh}. We list the $p$-value from a two sample KS-test in each panel. }
	\label{fig:illustris_outliers_ssfr}
	
\end{figure*}

\begin{figure*}
	\centering
	\tiny
\includegraphics[scale=0.40, trim=0.0cm 0.0cm 0.0cm 0.0cm,clip=true]{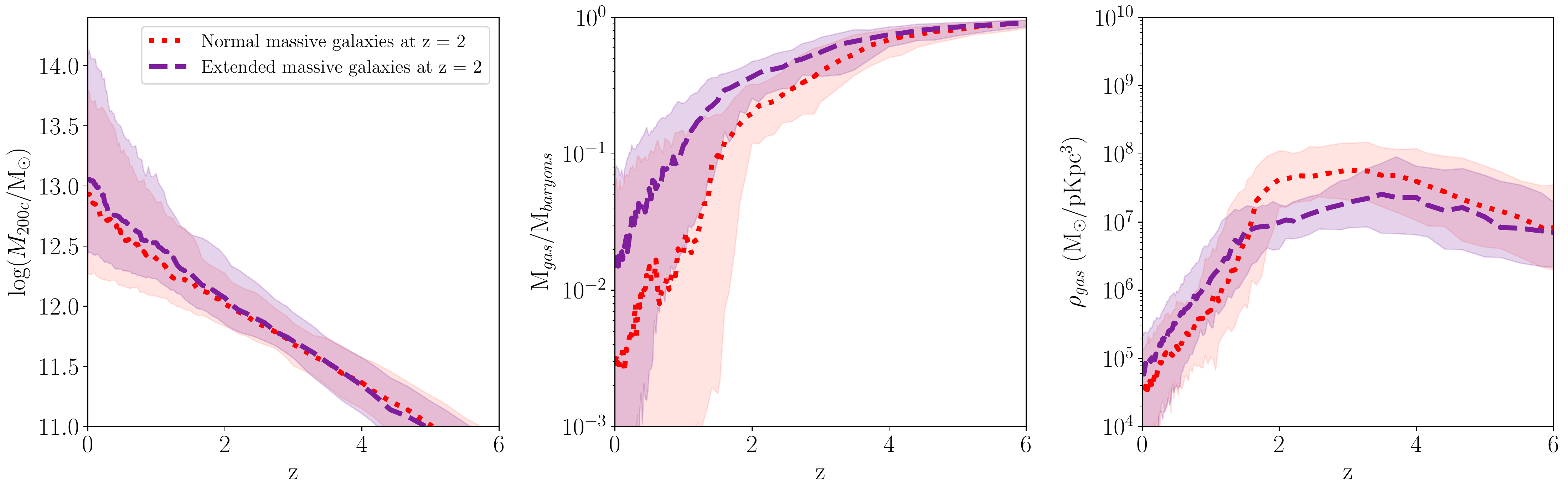}
	\caption{Median and 16$^{\rm th}-$84$^{\rm th}$ percentiles of halo mass (left panel), gas fraction (middle), and gas density (right panel) of the $z=2$-selected extended (purple) and normal (red) massive TNG100 galaxies, as a function of redshift. Both normal and extended massive galaxies occupy similar mass halos, but the extended massive galaxies  consistently have 2-10 times higher gas fraction compared to the normal massive galaxies after $z<4$ measured within twice the \rhalfmass. The normal massive galaxies have nearly five times higher gas density than the extended massive galaxies between $z=2-4$. }
	\label{fig:illustris_outliers_gas}
	
\end{figure*}

\section{Results} \label{sec:results}

\subsection{Star formation histories}

Figure \ref{fig:illustris_outliers_sfh} shows the stellar mass buildup and star formation histories of the extended and normal massive galaxies.  By selection, both extended and normal massive galaxies have similar sSFR at $z=2$. Even at $z>2.0$, the sSFR distributions of the extended and normal massive galaxies remains the same (KS-test $p>0.2$), suggesting both populations build up their stellar mass at a similar rate till $z=2.0$. The stellar mass distribution of the two populations remains statistically similar till $z=2$, although by $z=0$ the extended massive galaxies have on average $0.15$\,dex higher stellar mass than the normal massive galaxies. 

The star formation in the extended massive galaxies quenches later than in the normal-size massive galaxies (Figures \ref{fig:illustris_outliers_sfh} \& \ref{fig:illustris_outliers_ssfr}). By $z=1.5$ more than 41\% of the normal massive galaxies are quenched, whereas only 17\% of the extended massive galaxies are quenched. To calculate the quenched fraction, we use the  redshift dependent sSFR cut by \cite{Tacchella2019}.  Even till $z=1$, only 36\% percent of the extended massive galaxies are quenched in contrast to the 69\% quenched fraction in the normal massive galaxies. However, by $z\sim0.3$ both normal and extended massive galaxies have a similar quenched fraction. Our measurements suggest that by using  only stellar size, we can predict when star formation will quench in massive galaxies. 

\begin{figure}
	\centering
	\tiny
	\includegraphics[scale=0.45, trim=0.0cm 0.0cm 0.0cm 0.0cm,clip=true]{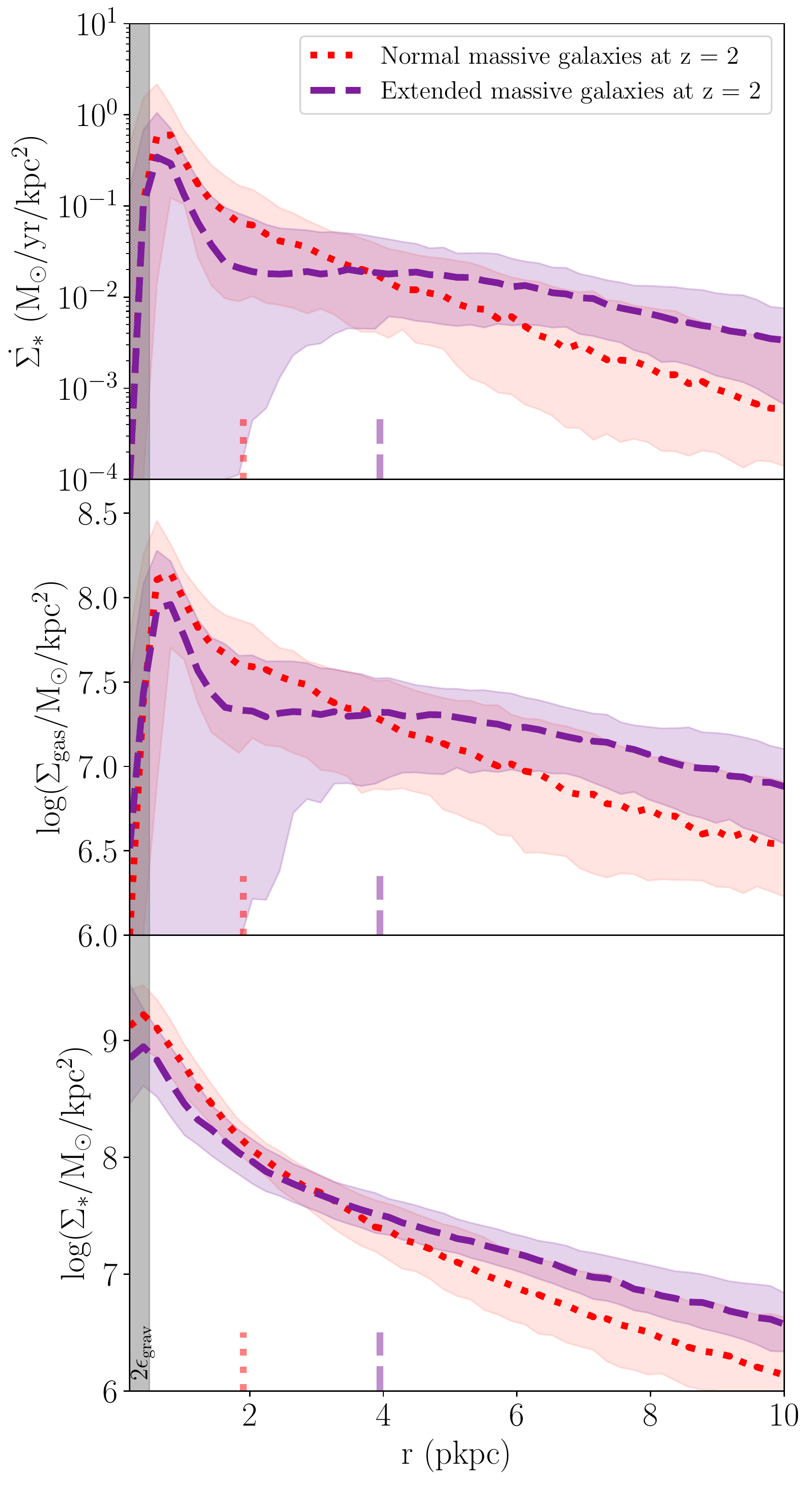}
	\caption{ Median radial star formation (top), gas mass (middle) and stellar mass (bottom) surface density  profiles for the normal (red dotted) and extended (purple dashed) massive galaxies at $z=2$. The \rhalfmass\ radius  for the respective population is given by the short vertical lines along the bottom edge of the each panel. The grey shaded region corresponds to the nominal resolution limit of TNG100 at $z=2$. The extended massive galaxies have lower SFR and gas mass surface density compared to the normal-size massive galaxies in their galactic centers. }
	\label{fig:illustris_radial_profile}
\end{figure}


%

\subsection{Quenching of star formation}

\subsubsection{Environment and gas density} 
Both observations and simulations show that galaxies living in more massive halos quench earlier \citep{Peng2010, Gabor2015, Wang2017}. However, the average halo mass of the extended massive galaxies remains within $1\sigma$ scatter to the average halo mass of the normal massive galaxies throughout the cosmic time (Figure \ref{fig:illustris_outliers_gas}: left panel). We do find that the median halo mass of the extended massive galaxies is  $\sim 0.08$\,dex higher than the normal massive galaxies, directly opposite to our expectations based on their delayed star formation quenching timescale. 

\cite{Martizzi2019} shows that at a fixed halo mass, galaxies residing in voids and sheets tend to  have lower stellar mass compared with the median of all galaxies, suggesting a clear role of the large-scale environment on the mass assembly history of galaxies. However, our preliminary analysis does not yield any significant evidence that the extended massive galaxies occupy a special place in the cosmic web compared with the normal massive galaxies. 

 Focusing on internal galaxy properties, the middle and right panels in Figure \ref{fig:illustris_outliers_gas} show how the fraction and the density of total gas (neutral and warm) within twice the \rhalfmass\ evolves with redshift. To get the average gas density, we divide the total gas mass within twice the stellar \rhalfmass\ with the total enclosed volume. The extended massive galaxies have systematically higher gas  fraction and gas density compared with the normal massive galaxies at $z<2$. It could be hypothesised that the higher gas fraction and gas density were responsible for the delayed onset of SF quenching  in the extended massive galaxies compared with the normal massive galaxies.

Interestingly, we find that the extended massive galaxies have lower gas density between $z=2-4$, i.e., just before the selection. The lower gas density in the extended massive galaxies could be responsible for their  relatively diffuse stellar core at $z=2$ and lower black hole feedback (See section \ref{sec:feedback}).

\begin{figure*}
	\centering
	\tiny
	\includegraphics[scale=0.45, trim=0.0cm 0.0cm 0.0cm 0.0cm,clip=true]{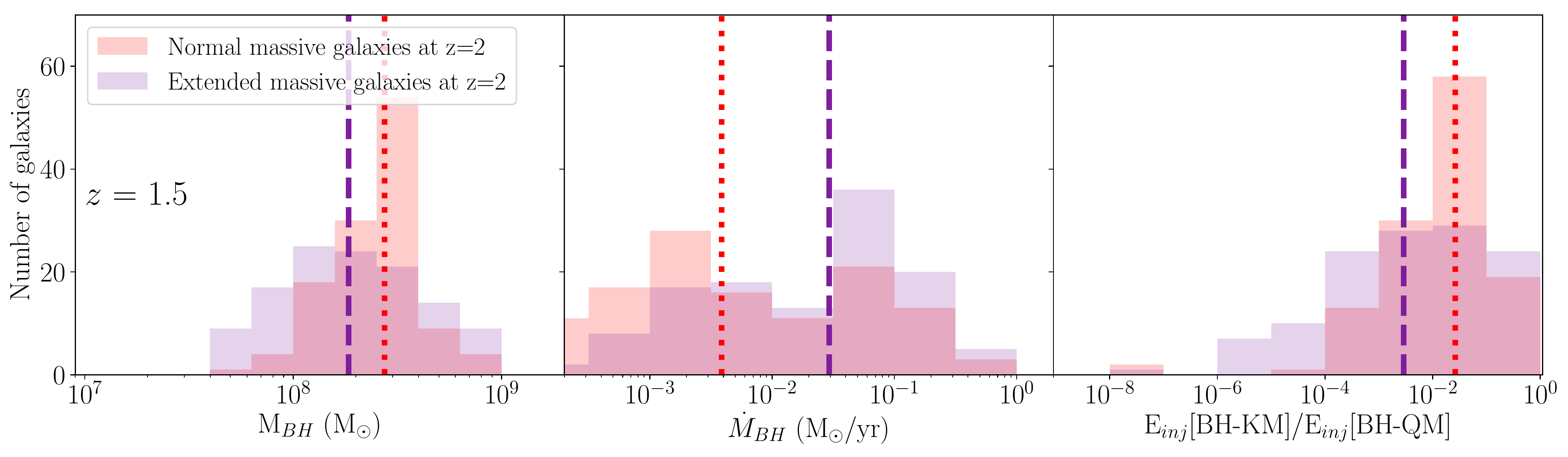}
	\includegraphics[scale=0.45, trim=0.0cm 0.0cm 0.0cm 0.0cm,clip=true]{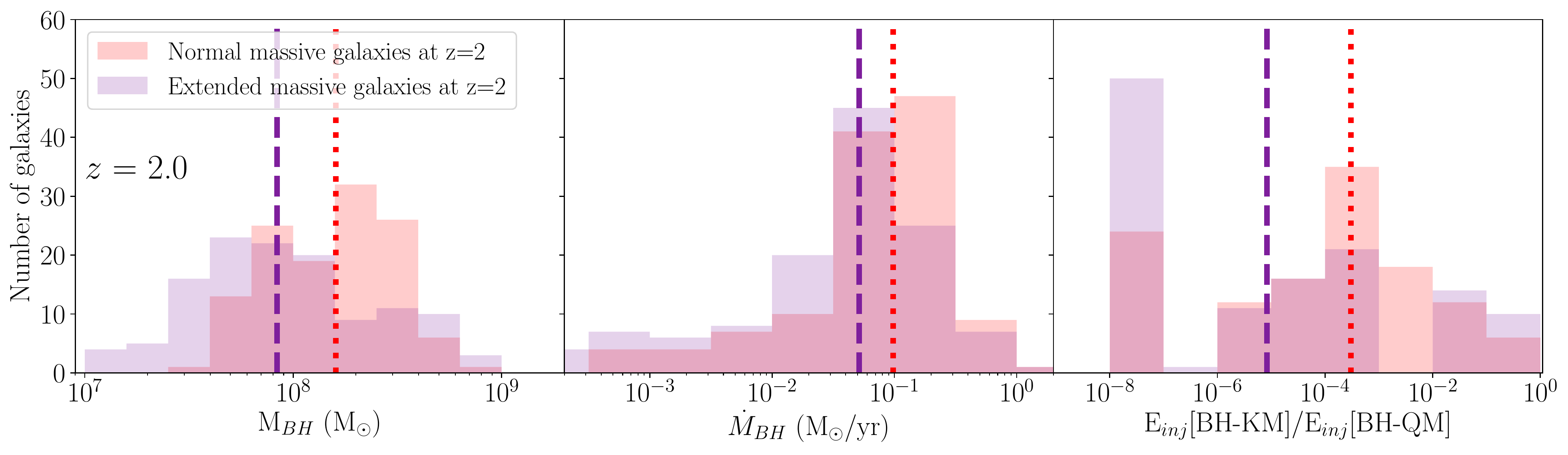}
	\includegraphics[scale=0.45, trim=0.0cm 0.0cm 0.0cm 0.0cm,clip=true]{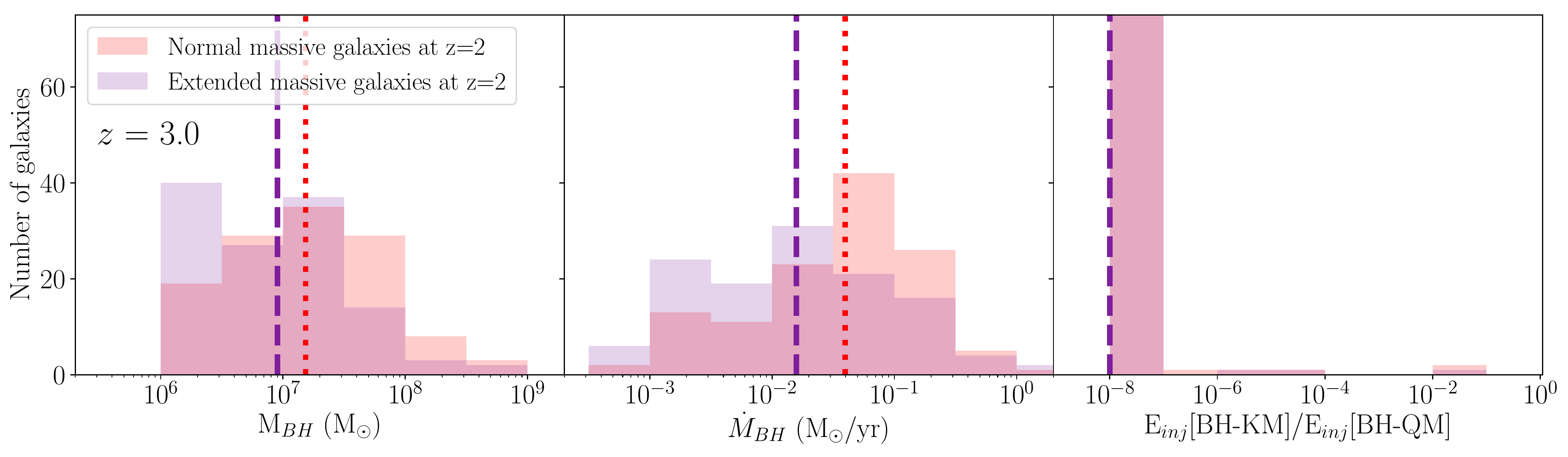}
	\caption{Distribution of the mass (left), accretion rate (middle) and ratio of cumulative energy injected in the kinetic-mode versus the thermal-mode (right) by the central supermassive  black hole in the normal (red) and extended (purple) massive galaxies at $z=1.5$ (top row), $2.0$ (middle row) and $3.0$ (bottom row). The dashed lines in each panel represent the median for the respective population. The normal massive galaxies have stronger black hole feedback than the extended massive galaxies at $z=2$. }
	\label{fig:illustris_bh}
\end{figure*}

\subsubsection{Central star formation and feedback from super massive black holes}\label{sec:feedback}

To understand the spatial variation of the SFR,  we estimate the radial star formation, gas mass and stellar mass surface density profiles in the extended and normal-size massive galaxies at the time of the selection (Figure \ref{fig:illustris_radial_profile}).  The extended massive galaxies have lower star formation and gas mass surface densities in their galactic center ($< 4$\,pkpc) in comparing with the normal massive galaxies. 

The star formation and gas mass density profiles remains almost flat between $2-10$\,pkpc for the extended massive galaxies. In contrast, the density profiles of the normal massive galaxies falls off rapidly beyond 2\,pkpc ($\sim$\rhalfmass). The extended galaxies have about 1\,dex higher star formation surface density at $r>4\,$pkpc. We hypothesise that the higher gas mass and star formation surface density in the galactic outskirts is responsible for the build-up of extended stellar disk in the extended massive galaxies. 

Observations in both the local and high redshift universe show that central stellar mass surface density correlates strongly with the onset of star formation quenching \citep{Fang2013, Barro2017, Mosleh2017}. The stellar mass surface density of the extended massive galaxies has a  shallower slope compared to the  normal massive galaxies. The extended galaxies have nearly 0.3\,dex lower stellar mass surface density in the central 1\,kpc than the normal massive galaxies. We suspect that the slow build-up of the central stellar core in the extended massive galaxies might be linked to their delayed star formation quenching timescale.

In massive galaxies, feedback from super massive black holes (SMBHs) is typically invoked for the  quenching of star formation \citep[see the review by][]{Fabian2012}. This is certainly the case in the IllustrisTNG model for galaxy formation \citep{Pillepich2017, Weinberger2017}.  In the IllustrisTNG model SMBH feedback operates in two different modes: the quasar-mode and the kinetic-mode. The quasar-mode occurs when the mass accretion rate is high, releasing enormous amount of thermal energy into the surrounding medium. The kinetic-mode occurs when the mass accretion rate is low and kinetic energy is deposited into the gas, producing jets and outflows or winds. SF quenching in particular is triggered by black hole driven winds at low-accretion rates \citep[see e.g.][]{Weinberger2017, Nelson2017a, Terrazas2019} because it determines the physical and thermodynamical properties of the gas within and around galaxies \citep[see][]{Truong2020, Davies2019, Zinger2020}.

Figure \ref{fig:illustris_bh}  shows that at $z=2$, the SMBHs in the extended massive galaxies are half as massive and accrete two times slower than the SMBHs in the normal-size massive galaxies. Already, at $z=3$, the normal-size massive galaxies host slightly more massive black holes than the more extended analogs. On the other hand, by $z=1.5$, the black hole accretion rate reverses and becomes almost 10 times higher in the extended massive galaxies probably due to the reduced supply of gas in the normal-size massive galaxies (Figure \ref{fig:illustris_outliers_gas}). 


\cite{Nelson2019,  Davies2019a, Terrezas2019, Zinger2020} show that in IllustrisTNG SMBHs quench star formation primarily by pushing the gas out from the central regions of galaxies via the kinetic-mode feedback.  The last column in Figure \ref{fig:illustris_bh} shows the ratio of the cumulative amount of energy injected by the SMBHs in the kinetic-mode versus the quasar-mode. In the extended massive galaxies, the ratio of energy injected in the kinetic-mode versus the quasar-mode is 10 times lower than the  normal-size massive galaxies at $z=2$. In IllustrisTNG,  kinetic-mode feedback becomes comparable to quasar-mode feedback once the SMBHs cross the transitional BH mass of  about $10^{8.3}$\msun\ \citep{Zinger2020}. Most extended massive galaxies have SMBH masses well below the transitional SMBH mass at $z=2$, and hence have relatively smaller fractions of energy injected in kinetic-mode.

Our analysis suggests that, at least in the case of IllustrisTNG galaxies, weaker AGN-driven kinetic feedback is responsible for the overall delayed quenching of star formation in the extended massive galaxies compared with the normal-size massive galaxies.  We suspect that the lower gas density in the galactic center  (Figure \ref{fig:illustris_radial_profile}) might be responsible for the slower black hole growth in the extended massive galaxies till $z=2$.

\begin{figure*}
	\centering
	\tiny
	
	\includegraphics[scale=0.55, trim=0.0cm 0.0cm 0.0cm 0.0cm,clip=true]{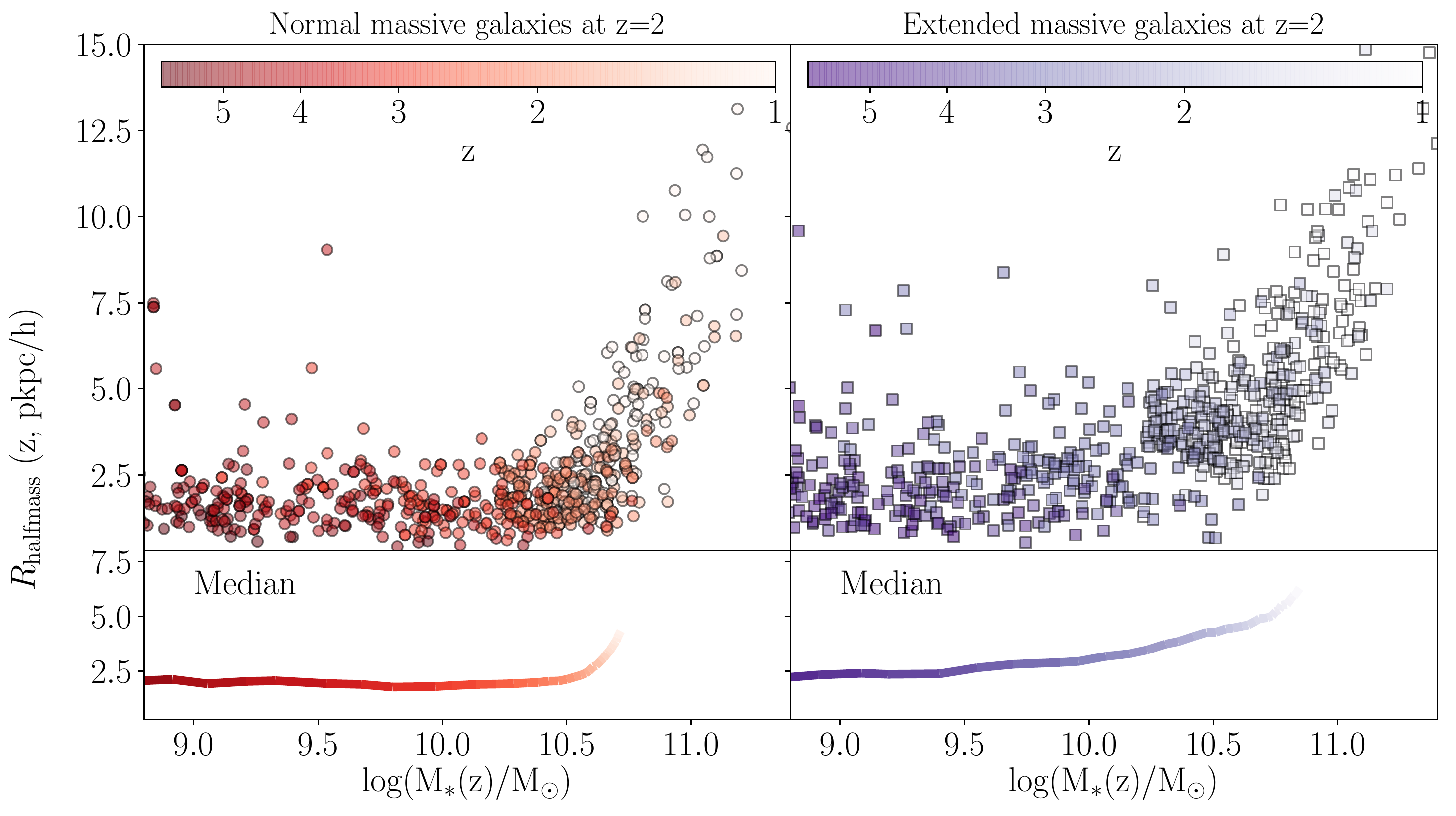}
	\caption{The evolutionary tracks corresponding to all 123 normal-size (left) and extended (right) massive TNG100 galaxies on the stellar size-mass plane.  Symbol colors indicate the redshift of the galaxies at that point. The colored solid curve in the bottom panel indicates the median tracks. The median stellar size of the normal-size massive galaxies does not change till $z\sim2$, whereas extended galaxies experience steady increase in their stellar size.  }
	\label{fig:illustris_size_sfr}
\end{figure*}

\subsection{Build up of the extended stellar disk}\label{sec:build_up}

The extended and normal-size galaxies follow a range of evolutionary pathways on stellar mass-size plane (Figure \ref{fig:illustris_size_sfr}). The median stellar size of the extended massive galaxies increases steadily as they build up their stellar mass between $z=2-4$. In contrast, the median stellar size-mass track of the normal-size massive galaxies is almost flat at least till $z=2.5$, i.e., on average normal-size massive galaxies build-up their stellar mass without a significant increase in their stellar size. Between $z=2-4$, the median stellar mass of the normal-size massive galaxies increases by 1 dex, while the median \rhalfmass\ radius remains roughly the same. Interestingly, \rhalfmass\ almost doubles within the same time period for the extended massive galaxies again with roughly 1\,dex increase in their stellar mass.

\begin{figure*}
		\centering
		\tiny
		\includegraphics[scale=0.55, trim=0.0cm 0.0cm 0.0cm 0.0cm,clip=true]{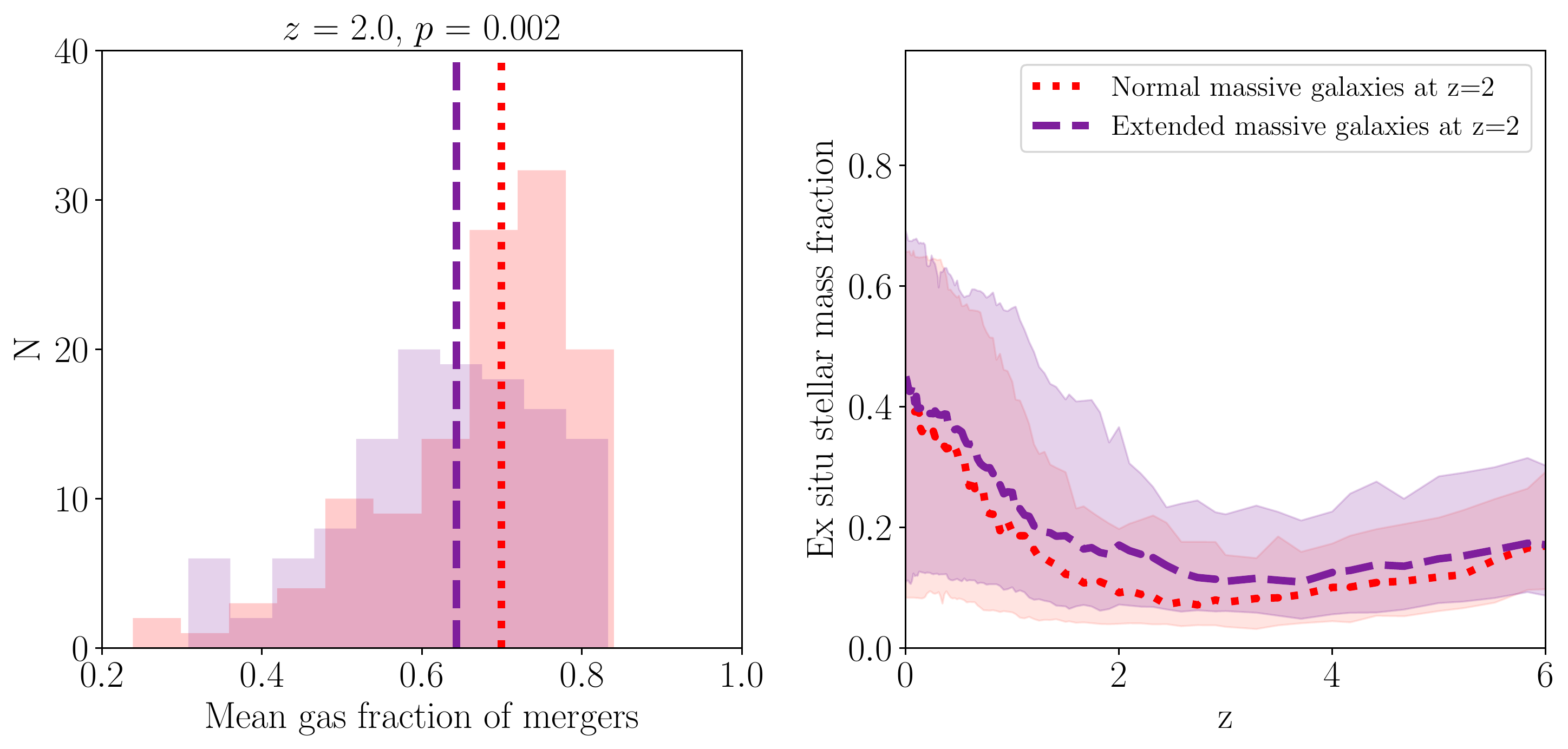}
		\caption{ The distribution of the mean cold fraction of the already merged galaxies at $z=2$ (left) and the evolution of the \exsitu\ stellar mass fraction (right) for the normal-size (red) and extended (purple) massive galaxies from TNG100. The dashed and dotted lines in each panel represents the median of the respective population.   }
		\label{fig:illustris_size_mergers}
	\end{figure*}

Using data from the \mosel\ survey in \cite{Gupta2020}, we find that massive galaxies at $z\sim2$ have higher integrated velocity dispersion compared to the counterpart galaxies at $z=3-4$. We conclude that build up of stellar mass  via \exsitu\ processes in massive galaxies after $z=2-3$ \citep{Rodriguez-Gomez2016, Gupta2020} increases the integrated velocity dispersion of gas. In our current analysis, we find that the median stellar size of normal-size massive galaxies only increases after $z<2$, again due to the stellar mass build-up via \exsitu\ processes (mostly dry merger, Figure \ref{fig:illustris_size_mergers}).

As in \cite{Gupta2020}, we use catalogs by \cite{Rodriguez-Gomez2016} to estimate the \exsitu\ stellar mass fraction, which is the fraction of stellar mass built via external processes such as mergers  and stripping from orbiting satellites.  The mean gas fraction of mergers in Figure \ref{fig:illustris_size_mergers} is taken from catalogs by \cite{Rodriguez-gomez2017} representing the average fraction of cold gas in already merged galaxies weighted by their stellar mass.  The median \exsitu\ fraction of the extended massive galaxies is $0.06-0.09$ higher than the normal-size massive galaxies, i.e., extended galaxies accrete a larger fraction of their stars. The accreted stars tend to reside at a larger galacto-centric distance \citep{Rodriguez-Gomez2016} and might be responsible for the higher  stellar surface density of the extended massive galaxies at large galacto-centric distance (Figure \ref{fig:illustris_radial_profile}). 

Moreover, we find that the mean gas fraction of mergers encountered by the extended massive galaxies by $z=2$ is lower by 10\% compared with the normal-size massive galaxies. The (mildly) lower gas fraction of mergers at $z=2$ coupled with the higher \exsitu\ fraction suggests that the extended massive galaxies have encountered relatively gas-poor mergers compared to the normal-size massive galaxies between $z=2-4$. Gas-rich mergers are excellent at funnelling the gas towards the galactic centers, driving central star formation and the growth of central supermassive black holes \citep{Hopkins2010,Medling2015, Wellons2015, Sparre2016, Hani2020}. Our analysis suggests that relatively drier mergers in the extended massive galaxies might be responsible for their lower central stellar mass density and slower SMBH growth.

\section{Discussion}\label{sec:discussion}

Observations of compact quiescent galaxies, `red-nuggets',  at $z\sim 2$ have triggered a fruitful debate over the onset of star formation quenching and the factors that may affect it in the early Universe \citep{Damjanov2009, Glazebrook2017, Lu2019}. In \cite{Gupta2020}, we found evidence that massive galaxies after $z\sim3$ buildup their stellar mass from \exsitu\ processes in both \mosel\ observations and IllustrisTNG simulations. We also observed a significant increase in the stellar sizes of the massive galaxies at $z\sim 2$. In this paper, we have analysed the TNG100 simulation to gain insights as to why the stellar sizes of some massive galaxies increase dramatically around $z=2$ and  how this affects their subsequent evolutionary history.

We find that, in the TNG100 simulation, the onset of star formation quenching in massive galaxies is connected to their stellar spatial distribution and extent. We find that $z=2$ galaxies with extended stellar distribution quench later than their relatively compact counterparts (Figure \ref{fig:illustris_outliers_sfh} \& \ref{fig:illustris_outliers_ssfr}). To minimise the bias at the time of selection, we have matched the stellar mass and SFR distributions of the extended and the normal-size massive galaxies at $z=2$. Our  findings are similar to those by \cite{Wu2018a}, who show that observed large quiescent galaxies at $z>1$ have younger stellar ages indicative of the relatively recent star formation quenching history.

	We hypothesise that relatively gas-poor mergers encountered by the extended massive galaxies at $z>2$ are responsible for their lower central stellar mass assembly by accretion and slower black hole growth (Figure \ref{fig:illustris_size_mergers}).  Gas-rich mergers are excellent at funneling  the gas towards the galactic centre triggering central star formation and the growth of the central SMBH \citep{Hopkins2010,Medling2015, Sparre2016, Hani2020}. Other physical mechanisms such as disk-instabilities introduced by the gas inflows and/or gravitation can also funnel the gas towards the galactic center triggering growth of central SMBH \citep{Dekel2014, Zolotov2015}. Relatively gas-poorer mergers between  $z=2-4$ could  explain the smaller black hole masses and lower gas density in the extended massive galaxies between $z=2-4$ (Figure \ref{fig:illustris_radial_profile} \& \ref{fig:illustris_bh}). 
	
	SMBHs in the extended  massive galaxies are almost half as massive and have injected a lower fraction of their energy in the kinetic-mode rather than in the quasar-mode compared with the  normal-size massive galaxies at $z=2$. In both the EAGLE and IllustrisTNG simulations, expulsion of  dense gas  has been shown to trigger the star formation quenching \citep[e.g.][]{Davies2019a}. \cite{Zinger2020} show that the  AGN-driven kinetic-mode feedback in IllustrisTNG  suppresses star formation by increasing the gas cooling timescales from 10-100\,Myr to 1-10\,Gyr. The lower fraction of kinetic energy injected by SMBHs in the extended massive galaxies due to their lower black hole masses would lead to weaker or less effective AGN-driven outflows, within the IllustrisTNG model, compared with the normal-size massive galaxies. This might be insufficient to trigger quenching of SF activity in the extended massive galaxies at $z=2$.

\cite{Habouzit2018} also find that compact massive galaxies in illustrisTNG simulations have higher star formation rates and stronger AGN feedback than their extended counterparts. Observationally, compact star-forming galaxies at $1<z<3$ have a higher fraction of AGNs  than the counterpart normal-size star-forming galaxies in line with our prediction from the IllustrisTNG simulations \citep{Barro2014, Rangel2014, Gu2020, Lu2020}. Our analysis suggests that weaker SMBH feedback is responsible for the delayed onset of  star-formation quenching in the extended massive galaxies. 
Future IFU surveys such as MUSE MAGPI survey (MAGPI collaboration in prep.) would be excellent to test our prediction that the onset of star formation quenching depends on the size of the stellar disk. 	


\section{Conclusions}\label{sec:conclusion}

Growth via dry-mergers has been shown to be responsible for the  dramatic increase in the scatter on stellar mass--stellar size plane at the massive end in \cite{Gupta2020} at $z=2$. In this paper,  we have used the TNG100 simulation of the IllustrisTNG project to gain insights on the connection between star formation activity and galaxy stellar sizes. In particular, we find that in the IllustrisTNG simulation $z=2$ massive galaxies with extended stellar distributions quench later than the normal-size galaxies with similar stellar masses and star formation rates at selection. 

 In particular, we have tracked the  evolutionary histories  of extended and normal-size massive galaxies (\logmstar$>10.2-11$) selected at $z=2$ with identical stellar mass and SFR distributions. By $z=1$, only 36\% of the extended massive galaxies have quenched, in contrast to more than 69\% quenched fraction in the normal-size massive galaxies (Figure 
\ref{fig:illustris_outliers_ssfr}). 

The stellar size of the extended massive galaxies almost doubles between $z=2-4$, with about 1\,dex increase in average stellar mass (Figure \ref{fig:illustris_size_sfr}). In contrast, the median stellar size of the normal-size massive galaxies does not change significantly between $z=2-4$, even if they experience similar increase in their stellar mass. At $z=2$, the extended massive galaxies have almost 0.3\,dex  lower stellar mass surface density in the central $1\,$kpc compared with the normal-size massive galaxies (Figure \ref{fig:illustris_radial_profile}). Stellar mass build-up via \exsitu\ processes leads to an increase in the sizes of both normal-size and extended massive galaxies after $z	<2$, as also shown in \cite{Gupta2020} using the \mosel\ survey \citep{Tran2020}.

We hypothesise that relatively gas-poorer mergers  experienced by the extended massive galaxies at $z>2$ are responsible for their lower central stellar mass density and SMBH mass. Our analysis shows that relatively lower SMBH mass and  weaker kinetic-mode  feedback from SMBHs are responsible for the delayed quenching of the star-formation activity in the extended massive galaxies (Figure \ref{fig:illustris_bh}). However, details of large-scale environment and merger histories need to be further investigated to understand why some massive galaxies experience gas-poor mergers and do not develop a central stellar core. 

\section*{Acknowledgements}
The authors thank the referee for their extremely valuable feedback and suggestions that significantly enhanced the quality of the paper. Parts of this research were supported by the Australian Research Council Centre of Excellence for All Sky Astrophysics in 3 Dimensions (ASTRO 3D), through project number CE170100013. SG is supported by the Simons Foundation through the Flatiron Institute. TNG100 was run on the HazelHen Cray XC40-system at the High Performance Computing Center Stuttgart as part of project GCS-ILLU of the Gauss Centre for Supercomputing (GCS). Ancillary and test runs of the IllustrisTNG project were also run on the Stampede supercomputer at TACC/XSEDE (allocation AST140063), at the Hydra and Draco supercomputers at the Max Planck Computing and Data Facility, and on the MIT/Harvard computing facilities supported by FAS and MIT MKI.

\bibliographystyle{aasjournal}

\end{document}